\documentclass[prl,twocolumn,amsmath,amssymb,superscriptaddress,longbibliography]{revtex4-2}

\usepackage{color,times}
\usepackage{bm}
\usepackage{dcolumn}
\usepackage{multirow}
\usepackage{amssymb,amsfonts,amsmath,graphicx}
\usepackage{epsfig}
\usepackage{xcolor}
\usepackage{SIunits}
\usepackage{braket}
\usepackage{epstopdf}
\usepackage{physics}
\usepackage{stmaryrd}
\usepackage{bbold}
\usepackage{amsmath}
\usepackage[utf8]{inputenc}
\usepackage{chngcntr}
\usepackage[english]{babel}
\usepackage[colorlinks=true, breaklinks=true, linkcolor=darkblue, citecolor=darkblue, urlcolor=darkblue]{hyperref}
\usepackage{textcomp}
\usepackage{soul}
\usepackage{makecell}
\usepackage{todonotes}

\definecolor{darkblue}{rgb}{0, 0, 0.8}
\definecolor{darkgreen}{rgb}{0, 0.5, 0}

\newcommand{\piot}{}

\counterwithout{equation}{section}
\addtocounter{equation}{0}

\begin{document}

\title{Enhancing a Many-body Dipolar Rydberg Tweezer Array with Arbitrary Local Controls}

\author{Guillaume~Bornet}
\affiliation{Universit\'e Paris-Saclay, Institut d'Optique Graduate School,\\
	CNRS, Laboratoire Charles Fabry, 91127 Palaiseau Cedex, France}

\author{Gabriel~Emperauger}
\affiliation{Universit\'e Paris-Saclay, Institut d'Optique Graduate School,\\
	CNRS, Laboratoire Charles Fabry, 91127 Palaiseau Cedex, France}

\author{Cheng~Chen}
\affiliation{Universit\'e Paris-Saclay, Institut d'Optique Graduate School,\\
	CNRS, Laboratoire Charles Fabry, 91127 Palaiseau Cedex, France}

\author{Francisco~Machado}
\affiliation{ITAMP, Harvard-Smithsonian Center for Astrophysics, Cambridge, Massachusetts, 02138, USA}
\affiliation{Department of Physics, Harvard University, Cambridge, Massachusetts 02138 USA}

\author{Sabrina~Chern}
\affiliation{Department of Physics, Harvard University, Cambridge, Massachusetts 02138 USA}

\author{Lucas~Leclerc}
\affiliation{Universit\'e Paris-Saclay, Institut d'Optique Graduate School,\\
	CNRS, Laboratoire Charles Fabry, 91127 Palaiseau Cedex, France}
\affiliation{PASQAL SAS, 7 rue L\'eonard de Vinci - 91300 Massy, Paris, France}

\author{Bastien~G\'ely}
\affiliation{Universit\'e Paris-Saclay, Institut d'Optique Graduate School,\\
	CNRS, Laboratoire Charles Fabry, 91127 Palaiseau Cedex, France}

\author{Daniel~Barredo}
\affiliation{Universit\'e Paris-Saclay, Institut d'Optique Graduate School,\\
	CNRS, Laboratoire Charles Fabry, 91127 Palaiseau Cedex, France}
\affiliation{Nanomaterials and Nanotechnology Research Center (CINN-CSIC), 
	Universidad de Oviedo (UO), Principado de Asturias, 33940 El Entrego, Spain}

\author{Thierry~Lahaye}
\affiliation{Universit\'e Paris-Saclay, Institut d'Optique Graduate School,\\
	CNRS, Laboratoire Charles Fabry, 91127 Palaiseau Cedex, France}

\author{Norman~Y.~Yao}
\affiliation{Department of Physics, Harvard University, Cambridge, Massachusetts 02138 USA}

\author{Antoine~Browaeys}
\affiliation{Universit\'e Paris-Saclay, Institut d'Optique Graduate School,\\
	CNRS, Laboratoire Charles Fabry, 91127 Palaiseau Cedex, France}

\begin{abstract}
We implement and characterize a protocol that enables arbitrary local controls in a dipolar atom array, where the degree of freedom is encoded in a pair of Rydberg states. 
Our approach relies on a combination of local addressing beams and global microwave fields. 
Using this method, we directly prepare two different types of  three-atom  entangled states, including a $W$-state and a state exhibiting finite chirality. 
We verify the nature of the underlying entanglement by performing quantum state tomography. 
Finally, leveraging our ability to measure multi-basis, multi-body observables, we explore the adiabatic preparation of low-energy states in a frustrated geometry consisting of a pair of triangular plaquettes. 
By using local addressing to 
tune the symmetry of the initial state, we demonstrate the ability to prepare correlated states distinguished only by correlations of their chirality (a fundamentally six-body observable). 
Our protocol is generic, allowing for rotations on arbitrary sub-groups of atoms within the array at arbitrary times during the experiment; this 
extends the scope of capabilities for quantum simulations of the dipolar XY model.
\end{abstract}

\date{\today}

\maketitle

The last decade has witnessed tremendous progress toward controllable 
many-body quantum systems~\cite{raimond:2001, leibfried:2003, ritsch:2013, gross:2017a, schafer:2020,heinrich:2021,daley:2022}.
This progress lies along two axes.
On the digital front, programmable interactions in small and intermediate scale systems can be compiled into arbitrary unitary evolution~\cite{arute:2019,iqbal:2023a,bluvstein:2024}.
On the analog front, a system's native interactions offer a scalable approach for realizing coherent many-body dynamics. 
This latter approach has emerged as a fruitful strategy for the quantum simulation of large-scale, strongly correlated many-body systems~\cite{bloch:2008,islam:2015a,bernien:2017a,deleseleuc:2019a,scholl:2021a}.
Combining the scalability of analog simulation with local controls inherent to the
digital approach promises the opportunity to explore a broader landscape of quantum phenomena.
In this quest for full many-body quantum control, various platforms ranging from neutral 
atoms~\cite{Bloch2012, Gross2017,Impertro2023} and trapped ions~\cite{Blatt2012, Monroe2021} 
to polar molecules~\cite{Zhou2011, Yan2013} and superconducting circuits~\cite{Houck2012,Kjaergaard2020},
have developed strategies to combine their  native interactions with high-fidelity local rotations.
This enhanced level of control has enabled the preparation of broader classes of initial states~\cite{fukuhara:2013,dumitrescu:2022}, 
the measurement of multi-basis observables~\cite{Roushan2017}, and even mid-evolution gates~\cite{zhang:2017}.
These advances 
enabled the integration of novel quantum information protocols with quantum simulators~\cite{Knill2008, Gambetta2012, Gaebler2012,Pagano_2020}.

Arrays of atoms coupled via Rydberg interactions have recently emerged as both promising quantum simulators~\cite{Browaeys2020, Chen2023} and information processors~\cite{Henriet2020, Morgado2021, Graham2022,Bluvstein2023}. 
Combining ground-state Raman manipulations~\cite{Yavuz2006,Jones2007} with the ability to address individual atoms has already allowed for the demonstration of local rotations in such systems~\cite{Isenhower2010,Xia2015,Birkl2007}. 
This is appropriate when the qubit is encoded, for example, in the hyperfine ground states of alkali atoms. 
However, when the qubit is encoded in a pair of Rydberg states, no analogous procedure has been realized. 
This challenge naturally arises for quantum simulations of the dipolar XY model~\cite{Whitlock2017,Leseleuc2017,Chen2023}, an important platform for the study of both correlated phases \cite{Chen2023} and quantum metrology~\cite{Bornet2023}.

Here, we address this challenge by 
demonstrating a general protocol implementing nearly-arbitrary local control in a dipolar Rydberg atom array.  
Our approach allows for the 
rotation of arbitrary classes of atoms and can be applied during any part of the experiment (i.e.~initialization, evolution and measurement). 
Our results are threefold. 
First, we benchmark our method by performing tomography on a three-atom $W$-state, demonstrating that it exhibits tripartite entanglement violating the Mermin-Bell inequality~\cite{Mermin1990}.
Second, by using local rotations to add phases to this $W$-state, we demonstrate the preparation of states exhibiting finite chirality~\cite{Wen1989}.
Both the measurement of this chirality, as well as the tomography of the system's density matrix,
require access to multi-basis, multi-body observables.
Finally, we extend our procedure to a frustrated six-atom system consisting of a pair of triangular plaquettes. 
By choosing different initial states, we adiabatically prepare states 
exhibiting both ferromagnetic and antiferromagnetic (six-body)
chiral-chiral correlation functions.
%

\begin{figure}
	\centering
	\includegraphics{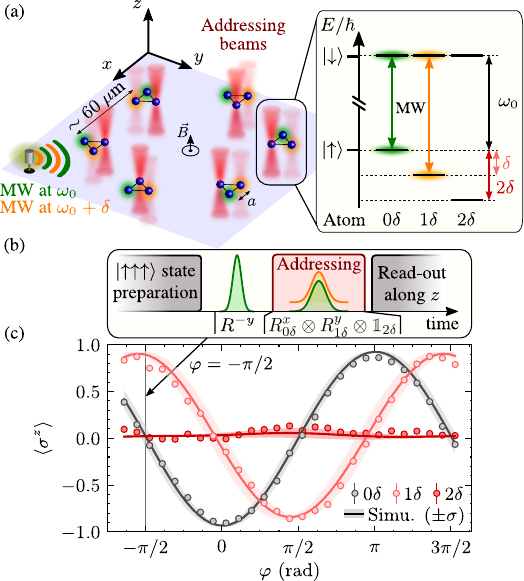}
	\caption{\textbf{Multi-basis measurements protocol.}
	(a)~ Experimental set-up. 
		The microwaves at a frequency $\omega_0$ ($\omega_0 + \delta$) 
		are on resonance with the $0\delta$ ($1\delta$) atom transitions and off-resonant with the others.
	(b)~Experimental sequence to measure the state of three atoms in the $y$, $z$ and $x$ basis. 		
	(c)~Average magnetization of each class during a Ramsey experiment. The experimental sequence for $\varphi=-\pi/2$ corresponds to the one in (b).
		Solid lines: simulations including experimental imperfections (see text).
		The shaded areas represent the standard deviation. }
	\label{fig:fig1}
        
\end{figure}

Our set-up \cite{Chen2023,Bornet2023,Chen2023a} consists of two-dimensional 
arrays of $^{87}\text{Rb}$ atoms trapped in optical tweezers. 
The atoms are arranged in groups of equilateral triangles [Fig.~\ref{fig:fig1}(a)]. 
We encode a qubit using two Rydberg states $\ket{\uparrow} = \ket{60S_{1/2},m_J=1/2}$ and  $\ket{\downarrow} = \ket{60P_{1/2},m_J=-1/2}$.
The atoms are coupled via dipolar interactions, described by the XY Hamiltonian:
\begin{equation}
\label{eq:H}
    H_{\rm XY}={J\over 2}
	\sum_{i < j}  \frac{a^3}{r_{ij}^3}  (\sigma^x_i \sigma^x_j + \sigma^y_i \sigma^y_j)~,
\end{equation}
with $r_{ij}$ being the distance between atom $i$ and $j$, $a=12.3(1)~\mu$m 
the lattice spacing, $J/a^3h= -0.82(1)~$MHz the interaction strength, and 
$\sigma_i^{x,y,z}$ the Pauli matrices acting on spin $i$. 
A $\sim 45~$G magnetic field, perpendicular to the atomic plane, 
defines the quantization axis and ensures isotropic interactions. 
At the beginning of each experimental sequence, the atoms are excited from their 
ground state to $\ket{\uparrow}$ using stimulated Raman adiabatic passage.

Our protocol to perform multi-basis measurements relies on the combination of microwave pulses and local light-shifts. 
The microwaves, tuned to the $\ket{\uparrow}-\ket{\downarrow}$ transition (at $\omega_0/(2\pi) \sim 16.7~$GHz), only allow for global rotations. To perform local rotations, we apply light-shifts on specific atoms using addressing beams generated by reflecting a $1013$\, nm laser 
on a spatial-light-modulator (SLM). This laser is blue-detuned with respect to the $6P_{3/2}-\ket{\uparrow}$ transition by $\Delta/(2\pi)\sim 400$\, MHz, resulting
in a light-shift $\delta\sim \Omega_{1013}^2/(4\Delta)$ for a Rabi frequency $\Omega_{1013}$ 
on an addressed atom~\cite{SM}. As illustrated in Fig.~\ref{fig:fig1}(a), 
the atoms are addressed with different intensities to 
produce different values of light-shifts, realizing the Hamiltonian 
$H_{z}=\sum_i \hbar \delta_i (1+\sigma^z_i)/2$, with $\delta_i=0\delta$ (atoms not addressed), 
$1\delta$ or $2\delta$ (with $\delta \approx 2\pi\times 23~$MHz). 
From now on, we refer to these classes of atoms as the $0\delta, 1\delta$ and  $2\delta$ atoms.
To perform local rotations on these three classes, 
we apply the addressing beams and send simultaneously two microwave 
pulses with frequencies $\omega_0$ and $\omega_0 + \delta$, 
resonant with the $0\delta $ and $1\delta$ atoms [see Fig.~\ref{fig:fig1}(a,b)]. 
This allows for arbitrary qubit rotations of the $0\delta$ and $1\delta$ atoms 
while the $2\delta$ atoms remain unaffected. 
By applying a global rotation prior to the local ones, as detailed below, 
we can now perform measurements in arbitrary bases on three classes of atoms at the same time: 
the choice of the measurement basis is set by the   
duration and phase of each microwave pulse with respect to a local oscillator at $\omega_0$.

As an example, Fig.~\ref{fig:fig1}(b) shows the experimental sequence used to measure the $0\delta$, $1\delta$ and $2\delta$ atoms along the $y$, $z$ and $x$ axis. The first microwave pulse applies a global $\pi/2$ rotation along the $-y$. We call $R^{-y}$ the corresponding rotation operator. Then, combining two microwave frequencies with the addressing, we apply the following local rotations $R^{x}_{0\delta} \otimes R^{y}_{1\delta} \otimes \mathbb{1}_{2\delta}$ with $R^{\bf u}_{n\delta}$ the operators corresponding to a $\pi/2$ 
rotation of the $n\delta$ atoms around the ${\bf u}$ axis. This full sequence is thus equivalent to the rotations 
$(R^{x}_{0\delta} \cdot R^{-y}_{0\delta})\otimes (R^{y}_{1\delta}\cdot R^{-y}_{1\delta}) \otimes R^{-y}_{2\delta}$. As, $R^{x}_{1\delta}\cdot R^{-y}_{1\delta}= R^{z}_{1\delta}\cdot R^{x}_{1\delta}$, and as we measure in the $z$-basis, the $z$ rotation has no effect on the measured probabilities. The sequence thus amounts to the rotation $  R^{x}_{0\delta} \otimes \mathbb{1}_{1\delta} \otimes R^{-y}_{2\delta}$ \footnote{Another, more natural, experimental protocol would have been to apply 3 microwave frequencies tuned on the 
$0\delta$, $1\delta$ and $2\delta$-atoms. However residual spatial inhomogeneities on the $2\delta$ light-shifts degraded 
the fidelities of the rotation in an early attempt.}.

We illustrate and benchmark the protocol above by performing a Ramsey experiment:
starting from all atoms in $\ket{\uparrow}$, we apply a first global rotation 
$R^{x\cos{\varphi} + y\sin{\varphi}}$, followed by the local rotations 
$R^{x}_{0\delta} \otimes R^{y}_{1\delta} \otimes \mathbb{1}_{2\delta}$ 
and finally read-out the states for various $\varphi$. 
Each experimental sequence is repeated $\sim 500$ times to compute the average magnetizations. 
We expect oscillations of the $0\delta$ and $1\delta$-atom magnetization that are out of phase by $\pi/2$.  
The $2\delta$-atom magnetization should remain constant at $0$. 
Figure~\ref{fig:fig1}(c) shows the experimental results. 
We attribute the finite contrast of the oscillations to experimental imperfections~\cite{SM}. 
To confirm this, we perform a Monte Carlo simulation 
including state preparation and measurement errors, finite Rydberg lifetime, 
interactions between atoms, and depumping and losses induced by the 
addressing~\cite{SM}. 
Taking into account all these experimentally calibrated mechanisms in the numerics 
yields good agreement with the data.

\begin{figure}
	\centering
	\includegraphics{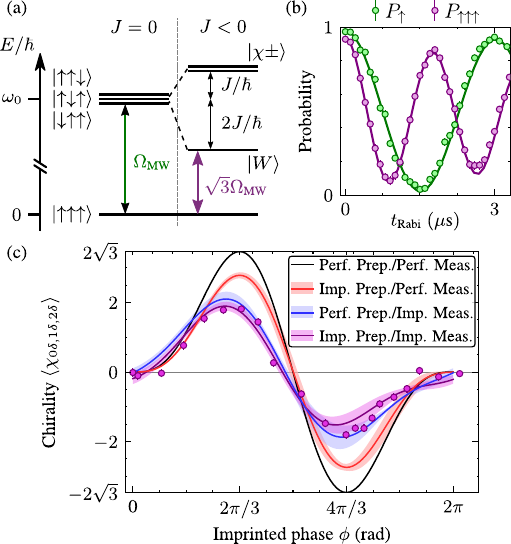}
	\caption{\textbf{$\ket{\chi(\phi)}$ state preparation.}
	(a)~Spectrum of 3 atoms in an equilateral triangle, interacting via the XY model. 
	(b)~Microwave Rabi oscillations. 
		Green: single-atom Rabi oscillations. 
		Purple: 3-atom Rabi oscillations with $P_{\uparrow\uparrow\uparrow}$ is the probability to measure all atoms in $\ket{\uparrow}$). 
		Solid curves: simulations including imperfections (see text).
	(c)~Chirality of $\ket{\chi(\phi)}$ as a function of $\phi$ (purple circles). 
		Purple, blue, red and black lines: Monte Carlo simulations including imperfections. 
		The shaded areas represent the standard deviation. }
	\label{fig:fig2}
\end{figure}

We now demonstrate how the local control introduced in our work enables the preparation and 
detection of complex, correlated states.
In particular, we investigate the entangled states of three atoms placed in an equilateral triangle and interaction  
via $H_{\rm XY}$. 
In this configuration, the interaction lifts the degeneracy between $\ket{\uparrow\uparrow\downarrow}$, 
$\ket{\uparrow\downarrow\uparrow}$ and $\ket{\downarrow\uparrow\uparrow}$, leading to three eigenstates 
$\ket{W} = \left( \ket{\uparrow\uparrow\downarrow} + \ket{\uparrow\downarrow\uparrow} + 
\ket{\downarrow\uparrow\uparrow} \right)/\sqrt{3}$ 
and $\ket{\chi^{\pm}} = ( \ket{\uparrow\uparrow\downarrow} + e^{\pm i\frac{2\pi}{3}} \ket{\uparrow\downarrow\uparrow} 
+ e^{\pm i\frac{4\pi}{3}}\ket{\downarrow\uparrow\uparrow} )/\sqrt{3}$ separated in frequency by $3J/\hbar$, 
as shown in Fig.~\ref{fig:fig2}(a).
Despite all states exhibiting homogeneous magnetization and two-point correlation functions, 
they can be distinguished through their \textit{chirality}.  
The chirality $\chi$ is a spin rotationally symmetric observable that breaks time reversal symmetry and is defined for three spins $i$, $j$ and $k$ by $\langle\chi_{ijk} \rangle= \langle({\bm \sigma}_i \times {\bm \sigma}_j) \cdot {\bm\sigma}_k\rangle$, 
with ${\bm \sigma}_i = \sigma_i^x{\bm x} + \sigma_i^y{\bm y} + \sigma_i^z{\bm z}$~\cite{Tsomokos2008}. 
For a product state, $\langle \chi \rangle$ is bounded by $\pm1$ and this limit can be overcome for entangled states. For $\ket{\chi^{\pm}}$, it reaches a maximum value of 
$\pm2\sqrt{3}$.

In order to prepare these states, we proceed as follows. 
Starting from all atoms in $\ket{\uparrow}$,
we apply a Gaussian microwave pulse at frequency $\omega_0 + 2J/\hbar\,$ to drive a 
direct transition from $\ket{\uparrow\uparrow\uparrow}$ to $\ket{W}$.
The Rabi frequency is collectively enhanced by a factor of $\sqrt{3}$, compared to the one measured for single atom Rabi oscillation experiment (see Fig.~\ref{fig:fig2}(b)).
These dynamics are well captured by numerical simulations that include 
all identified imperfections (see \cite{SM}).  
Finally, we turn on the addressing light for a duration $t_{\text{phase}}$ 
to imprint a phase $0\phi$, $1\phi$ and $2\phi$ on the $0\delta$, $1\delta$ and $2\delta$ atoms, with $\phi(t_{\text{phase}}) = \int_{0}^{t_{\text{phase}}} \delta(t) \,{\rm d}t$, 
thus preparing $\ket{\chi(\phi)} = \left( \ket{\uparrow\uparrow\downarrow} + 
e^{ i\phi}\ket{\uparrow\downarrow\uparrow} + e^{i2\phi}\ket{\downarrow\uparrow\uparrow} \right)/\sqrt{3}$.

To measure the chirality, we first note that it can be written as the sum of six terms corresponding to 
the different permutations of $\{x,y,z\}$:~$\langle \chi_{0\delta,1\delta,2\delta} \rangle = \langle \sigma^x_{0\delta}\sigma^y_{1\delta}\sigma^z_{2\delta}\rangle + \langle \sigma^y_{0\delta}\sigma^z_{1\delta}\sigma^x_{2\delta}\rangle + \langle \sigma^z_{0\delta}\sigma^x_{1\delta}\sigma^y_{2\delta}\rangle - \langle \sigma^y_{0\delta}\sigma^x_{1\delta}\sigma^z_{2\delta}\rangle - \langle \sigma^x_{0\delta}\sigma^z_{1\delta}\sigma^y_{2\delta}\rangle -\langle \sigma^z_{0\delta}\sigma^y_{1\delta}\sigma^x_{2\delta}\rangle$.
For each value of $\phi$, we measure each set of bases to compute the 
total chirality of $\ket{\chi(\phi)}$, 
similarly to previous work using superconducting qubits ~\cite{Roushan2017}.
Figure~\ref{fig:fig2}(c) shows the results (purple circles) as a function of $\phi$, 
together with the theoretical expectations. The amplitude is reduced due to experimental imperfections. 
Simulating each step of the sequence while accounting for these imperfections
\cite{SM} leads to a better agreement between theory and experiment. 
From a simulation of the measurement sequence (including local rotations and the readout step) 
assuming a perfect state preparation [blue curve in Fig.~\ref{fig:fig2}(c)], 
we find that the main limitations of the chirality measurement are the 
imperfections during the measurement sequence.

\begin{figure}
	\centering
	\includegraphics[]{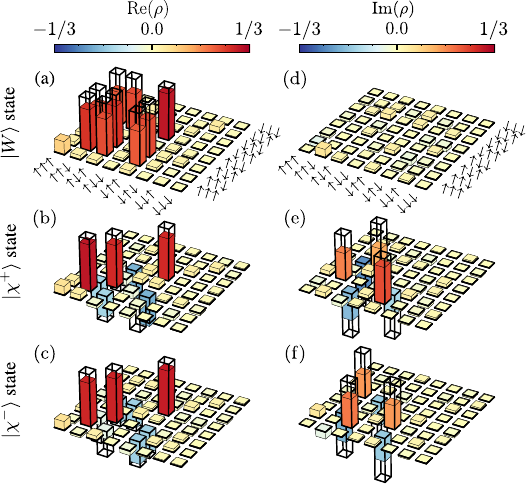}
	\caption{\textbf{State tomography.} Real (a,b,c) and imaginary (d,e,f) parts of the 
	reconstructed density matrix for the $\ket{W}$, $\ket{\chi^+}$ and $\ket{\chi^-}$ states. 
	The transparent bars represent the expectation values for perfect states. }
	\label{fig:fig3}
\end{figure}

We now exploit our ability to apply arbitrary local rotations to perform quantum state tomography of
$\ket{W}$ and $\ket{\chi^\pm}$ and reconstruct their density matrix. 
To do so, we measure the state of each class of atoms in the $x$, $y$ and $z$ bases, corresponding to $3^3=27$ 
different measurements~\cite{SM}, from which we compute the relevant correlation functions, 
as well as extract the density matrix (using a maximum-likelihood reconstruction~\cite{SM}).  
Figure~\ref{fig:fig3} shows, for one triangle, the real and imaginary parts of the density matrices 
$\rho$ of the three states $\ket{W}$, $\ket{\chi^+}$ and $\ket{\chi^-}$.
From them, we compute fidelities $F = \expval{\rho}{\psi}$ of $0.74(1)$, $0.71(1)$ and $0.68(1)$ 
[$0.80(1)$, $0.78(1)$ and $0.74(1)$ when correcting for detection errors]. 
They are all above $2/3$, revealing genuine three-partite entanglement~\cite{Acin2001, Guehne2003, Neeley2010}. 
In addition, the produced $W$-state violates the Mermin-Bell inequality: 
$S  = \lvert \langle \sigma^z_{0\delta}\sigma^z_{1\delta}\sigma^z_{2\delta}\rangle - \langle \sigma^x_{0\delta}\sigma^x_{1\delta}\sigma^z_{2\delta}\rangle - \langle \sigma^z_{0\delta}\sigma^x_{1\delta}\sigma^z_{2\delta}\rangle - \langle \sigma^z_{0\delta}\sigma^z_{1\delta}\sigma^x_{2\delta}\rangle \lvert \leq 2$ as we
measure $S_{\rm exp} = 2.083(26)$~\cite{Mermin1990}. 
Much like in the more conventional Bell-state case, this violation rules out a hidden-variable model for the measured correlations.

\begin{figure}[t]
	\centering
 \includegraphics[width=3.2in]{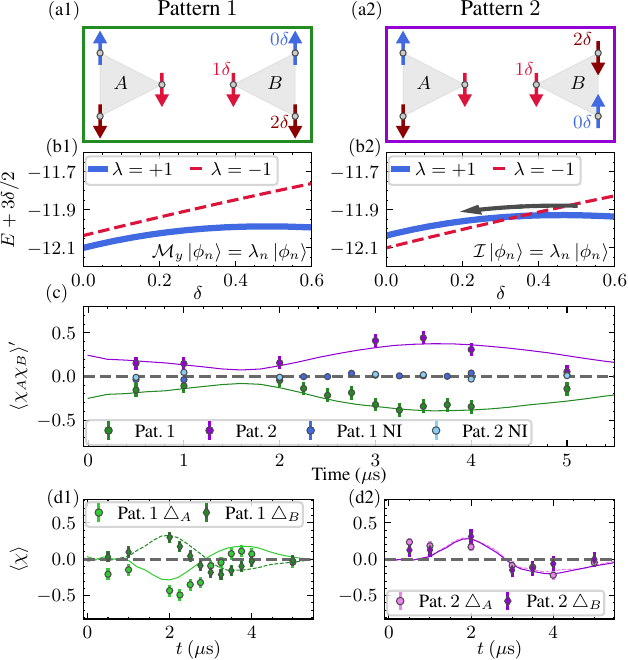}
  \caption{{\bf Adiabatic preparation of low energy states}
        {\bf (a)} Light shift patterns $(0,1\delta, 2\delta)$ and associated initial states.
        Pattern 1 respects mirror symmetry along the $y$-direction $M_y$ whereas pattern 2 respects inversion symmetry $\mathcal{I}$.
        {\bf (b)} 
        Energy spectrum during the adiabatic ramp for pattern 1[2].
        {\bf (c)} Chiral-chiral correlations $\langle \chi_A\chi_B\rangle^{\rm '}$ for pattern 1 (green), pattern 2 (purple) and non interacting triangles (blue). The early time ($t\lesssim 1.5\mu s$) observation of non-zero $\langle \chi_A\chi_B\rangle^{\rm '}$ is due to a necessary waiting period before measurement, during which the system undergoes additional dynamics (see~\cite{SM}).  
        {\bf (d)}
		Chirality for triangles A and B under the two patterns. 
        }
        \label{fig:fig4}
\end{figure}

Having leveraged our local control to prepare and probe entangled states, we now demonstrate the power of this toolset in a quantum simulation experiment.
Using a frustrated geometry consisting of a pair of triangular plaquettes [Fig.~\ref{fig:fig4}(a1)], we attempt to adiabatically prepare low-energy states of the antiferromagnetic dipolar XY model \footnote{Strictly speaking, since in Eq.~(\ref{eq:H}) we have ferromagnetic couplings $J<0$, we explore the low-energy properties of the antiferromagnetic model ($J>0$) by preparing the highest-energy state(s) of our quasi-isolated system.}.
Owing to time-reversal symmetry, all states in the spectrum exhibit zero chirality, $\langle\chi\rangle =0$. 
However, exact diagonalization demonstrates that, the two lowest-energy states exhibit large, but opposite, chiral-chiral correlations.
To illustrate this feature, we prepare both the ground and first excited states, by carefully choosing an appropriate pattern of local light shifts. 

Our protocol proceeds as follows~\cite{Chen2023}:
after initializing all the atoms in $\ket{\downarrow}$, we turn on
a pattern of local light shifts~[Fig.~\ref{fig:fig4}(a)]; we then apply a microwave pulse to rotate the non-addressed atoms to $\ket{\uparrow}$.
This prepares a product state which is the lowest energy state of $H_z$. Starting with $\delta\gg |J|$, we then reduce the light-shift as 
$\delta(t) = \delta_0 e^{-t/\tau}$ [with $\tau = 0.55~\mu$s and $\delta_0/2\pi = 23(46)~$MHz for the $1\delta(2\delta)$-atoms], thus quasi-adiabatically connecting the initial Hamiltonian ($\approx H_z$) to the final one $H_{\rm XY}$.

In general, such an adiabatic protocol is expected to prepare the ground state of the final Hamiltonian, regardless of the details of the ramp.
This expectation fails when the system's ground state exhibits a level crossing, which requires either fine-tuning or some underlying symmetry.
Utilizing our ability to shape the addressing light, we thus consider two different patterns exhibiting distinct symmetries: pattern 1 respects a mirror symmetry $M_y$ along the y-axis  [Fig.~\ref{fig:fig4}(a1)], while pattern 2 respects inversion symmetry $\mathcal{I}$ [Fig.~\ref{fig:fig4}(a2)].
For the first pattern, both initial and ground states live in the same symmetry sector of $M_y$ and thus are adiabatically connected [Fig.~\ref{fig:fig4}(b1)].
We thus expect to prepare the ground state, leading to the observation of \emph{anti-ferromagnetic} chiral-chiral correlations.
By contrast, for the second pattern, the initial and ground states live in \emph{different} symmetry sectors of $\mathcal{I}$ and thus cannot be adiabatically connected  [Fig.~\ref{fig:fig4}(b2)]. We thus expect to prepare the first excited state, which exhibits \emph{ferromagnetic} chiral-chiral correlations.

We experimentally explore this difference using the multi-basis measurement protocol described above. It not only allows measuring the chirality $\langle \chi (t)\rangle$ of a single triangle, but it also enables the measurement of six-body correlation functions, from which the two-triangle chiral-chiral correlations, $\langle \chi_A\chi_B\rangle$, can be extracted.
In principle, the full reconstruction of $\langle \chi_A\chi_B\rangle$ requires the measurement of 36 different terms. 
However, numerical simulations indicate that a smaller subset of six terms is sufficient to faithfully capture the system's correlations~\cite{SM}.
We can, therefore, use the same addressing pattern (and thus only a single SLM~\cite{SM}) for both the adiabatic ramp and the multi-basis measurement.
More specifically, we measure: 
\begin{align}
\label{eq:chi-chiCorr}
\langle \chi_A \chi_B \rangle'  =  \eta\hspace{-6mm}\sum_{\substack{a,b,c ~\in~\text{perm}(x,y,z)}} \hspace{-4mm}&\begin{array}{l}
     \langle \sigma^a_{0\delta}\sigma^b_{1\delta}\sigma^c_{2\delta}\tilde{\sigma}^a_{0\delta}\tilde{\sigma}^b_{1\delta}\tilde{\sigma}^c_{2\delta}\rangle  \\ \\[-3mm]
     \quad \left.-\langle \sigma^a_{0\delta}\sigma^b_{1\delta}\sigma^c_{2\delta} \rangle \langle \tilde{\sigma}^a_{0\delta}\tilde{\sigma}^b_{1\delta}\tilde{\sigma}^c_{2\delta} \rangle \right.
\end{array}
\end{align}
where $\sigma[\tilde{\sigma}]$ refers to spins in triangle A[B] and $\eta = \pm 1$ is set by the relative handedness of the two three-spin measurement patterns: $\eta=-1$ for pattern 1 and $\eta=1$ for pattern 2~\cite{SM}.

We begin by studying the quasi-adiabatic ramp using the pattern depicted in Fig.~\ref{fig:fig4}(a1). 
Focusing on the chiral-chiral correlation, we observe the development of strong \emph{anti-ferromagnetic} $\langle \chi_A\chi_B \rangle^{\rm '}$ correlations that persist to late times [Fig.~\ref{fig:fig4}(c), green].
This observation is consistent with a preparation yielding more than $50\%$ population in the ground state \cite{SM}.
By contrast, when considering the second pattern [Fig.~\ref{fig:fig4}(a2)], the dynamics exhibit similar features but with \emph{opposite} sign.
The presence of equally strong \emph{ferromagnetic} $\langle \chi_A\chi_B \rangle^{\rm '}$ correlations is consistent with an equally large population in the first excited state of the system.
To demonstrate that our observations indeed arise from the dipolar interactions between the two triangles, we also measure $\langle \chi_A\chi_B \rangle^{\rm '}$ for non-interacting triangles separated by $\approx 72~\mathrm{\mu m}$.
In this case, neither patterns lead to significant correlations [Fig.~\ref{fig:fig4}(c), blue].

Finally, we discuss two important sources of imperfections in our protocol.
First, although $\tau$ was chosen to be much longer than the timescale of the system ($1/J$), residual diabatic errors manifest themselves in a small chirality value $\langle \chi(t)\rangle$ [Fig.~\ref{fig:fig4}(d)].
Second, there are fluctuations in the positions of the atoms owing to their initial position and velocity uncertainty upon the release from the tweezers.
As a result, for each repetition of the experiment, the atoms experience slightly different time-dependent interactions, that ultimately lead to the damping of the chirality oscillation and to the decay of the chiral-chiral correlations~\cite{SM}.

In conclusion, we have demonstrated a new tool combining global microwaves and local light-shifts to enable local control of qubits encoded in Rydberg levels. 
Our protocol is generic and can be extended to an arbitrary number of classes of atoms. The agreement between experiments and simulations highlights our good understanding of error in our system---a crucial ingredient for further improvements.

More broadly, this work opens the doors to a number of intriguing directions.
First, the measurement of multi-body correlation functions can capture the intricate correlations that characterize complex phases of matter such as time reversal symmetry breaking and topological order~\cite{Wen1989}.
Second, the ability to measure along arbitrary bases enables the implementation of novel certification protocols~\cite{Notarnicola2023}.
Finally, by interspersing unitary rotations with analog quantum simulation, one can study multi-time correlation functions as well as more varied dynamical protocols~\cite{Kokail2019,Roushan2017,Brydges2019}.

\begin{acknowledgments}
	This work is supported by
	the Agence Nationale de la Recherche (ANR-22-PETQ-0004 France 2030, project QuBitAF), 
	and the European Research Council (Advanced grant No. 101018511-ATARAXIA), and 
	the Horizon Europe programme HORIZON-CL4- 2022-QUANTUM-02-SGA (project 101113690 (PASQuanS2.1), and the US Army Research Office (W911NF-21-1-0262), and the AFOSR MURI program  (W911NF-20-1-0136).
	DB acknowledges support from MCIN/AEI/10.13039/501100011033 
	(RYC2018-025348-I, PID2020-119667GA-I00, and European Union NextGenerationEU PRTR-C17.I1)
	FM acknowledges support from the NSF through a grant for ITAMP at Harvard University. 
    SC acknowledges support from the National Science Foundation Graduate Research Fellowship under Grant No. DGE 2140743.
\end{acknowledgments}

\bibliography{references_XYZ_paper, Exported_Items}

\clearpage

\setcounter{figure}{0}
\renewcommand\thefigure{S\arabic{figure}} 

\begin{center}
{\bf Supplemental Material}
\end{center}


\section{Experimental methods}\label{SM:Exp_details}

The implementation of the dipolar XY Hamiltonian relies on the $^{87}\text{Rb}$ 
Rydberg-atom tweezer array platform described in previous works~\cite{Chen2023,Bornet2023,Chen2023a}. 
The tweezer array is created by diffracting a 820-nm laser beam on a first Spatial Light Modulator (SLM)
and focusing it by a $\text{NA}=0.5$ aspherical lens~\cite{Nogrette2014}. 
The phase pattern imprinted on the beam governs the geometry of the array. 
This pattern cannot be changed during the experimental sequence. 

We encode our qubit states as $\ket{\uparrow} = \ket{60S_{1/2}, m_J = 1/2}$ 
and $\ket{\downarrow} = \ket{60P_{3/2}, m_J = -1/2}$ and use microwaves at $\omega_0/(2\pi)\approx16.7$\,GHz 
to manipulate them. A $45$-G quantization magnetic field perpendicular to the array ensures 
isotropic XY interactions and shifts away from the $\ket{\uparrow}-\ket{\downarrow}$ 
transition the irrelevant Zeeman sublevels. 

The microwaves used to manipulate the qubit states are generated using a vector signal generator (R\&S\textsuperscript{\textregistered}SMM100A) plugged to an antenna placed outside the vacuum chamber.

\subsection{Experimental sequence}\label{SubSM:sequence}

\begin{figure*}
	\centering
	\includegraphics{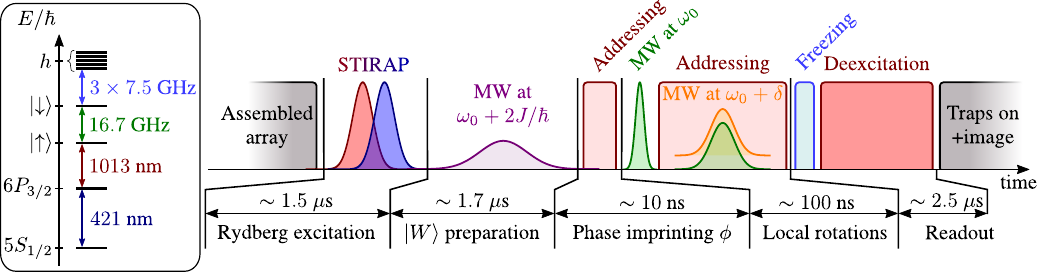}
	\caption{\textbf{Detailed experimental sequence. }
	(Left) Atomic levels used in this work. 
	(Right) Experimental sequence used to prepare a $\ket{\chi(\phi)}$ state, 
	from the Rydberg excitation step to the readout.}
	\label{fig:fig_SM_sequence}
\end{figure*}

\emph{Assembled array of Rydberg atoms} -- 
Figure~\ref{fig:fig_SM_sequence} 
shows the detailed experimental sequence used to prepare the $\ket{\chi^\pm}$ 
states and measure the chirality. 
After randomly loading atoms into optical tweezers of 1-mK depth (with a typical filling fraction of $60\%$), 
we assemble the array~\cite{Barredo2016}. We then cool the atoms to a temperature of $10\,\mu$K 
using Raman sideband cooling and optically pump them to $\ket{g}=\ket{5S_{1/2}, F = 2, m_F = 2}$. 
Following this, we adiabatically ramp down the power of the trapping light to reduce 
the tweezer depth by a factor $\sim 2$. Then, we switch off the tweeezers and 
excite the atoms to the Rydberg state $\ket{\uparrow}$. 
The excitation is performed by  a two-photon stimulated Raman adiabatic 
passage (STIRAP) with 421-nm and 1013-nm lasers.

\emph{Preparation of the state $\ket{W}$} -- 
To generate the $\ket{W}$ state, we apply a  microwave pulse with gaussian temporal profile
$\Omega(t) = \Omega_{\max}e^{-\pi \left( t/t_{\text{Rabi}} \right)^2}$, 
detuned by $2J$ from the resonance, with a Rabi frequency  $\Omega_{\max}/2\pi = 0.33~$MHz 
and a pulse duration of $t_{\text{Rabi}}=0.950~$ns.

\emph{Preparation of the state $\ket{\chi(\phi)}$} -- 
We then turn on the addressing for a time $t_{\text{phase}}$ of a few nanoseconds to imprint a phase $0\phi$, $1\phi$, and $2\phi$ on the 
$0\delta$, $1\delta$ and $2\delta$ atoms respectively (see next section for more details about the addressing set-up). 
Experimentally, we use a free-space Electro-Optic Modulator (EOM) to control this pulse. 
Due to transient effects the imprinted phase is not linear with $t_{\text{phase}}$ and 
rather reads $\phi(t_{\text{phase}}) = \int_{0}^{t_{\text{phase}}} \delta(t) \,dt$. 
We calibrated $\phi(t_{\text{phase}})$ using a dedicated experiment 
(see next sections).

\emph{Local rotations sequence} -- 
Once a state $\ket{\chi(\phi)}$ is prepared, we apply local rotations on the 
$0\delta$, $1\delta$ and $2\delta$ atoms to measure them in different basis. 
As explain in the main text, we first send a first  microwave pulse (Gaussian temporal profile)
with a frequency $\omega_0$ and an amplitude $\Omega_{\max}/2\pi = 19.23~$MHz 
(a $\pi/2$ rotation corresponds to $t_{\text{Rabi}}=13~$ns) 
to perform a global rotation on all the atoms. We then turn of the addressing light for $80~$ns 
and during this time send two microwave Gaussian pulses at frequency 
$\omega_0$ and $\omega_0 + \delta$ and with a same amplitude of  $\Omega_{\max}/2\pi = 5.43~$MHz 
(a $\pi/2$ rotation corresponds to $t_{\text{Rabi}}=46~$ns). 
We independently control the pulse duration and phase of each microwave pulse, 
allowing to perform any arbitrary rotations.

\emph{Readout} -- 
To avoid residual interactions between the atoms during the readout procedure 
we first apply a microwave pulse at $7.5~$GHz transferring the atomic population 
from $\ket{\downarrow}$ to the $n=58$ hydrogenic manifold $(h)$ via a three-photon transition. 
The atoms in $(h)$ do not interact with those remaining in $\ket{\uparrow}$. 
The interaction dynamics is then frozen. The next step consists in a deexcitation 
pulse performed by applying a $2.5~\mu$s laser pulse on resonance with the 
transition between $\ket{\uparrow}$ and the short-lived intermediate state $(6P_{3/2})$ 
from which the atoms decay back to ground state $5S_{1/2}$. 
Finally we ramp back on the trapping lights recapturing only the atoms in the ground state while the ones in $(h)$ are expelled from the traps. 
When then turn on the fluorescence beams, and image the recaptured atoms. 
Thus we map the $\ket{\uparrow}$ and $\ket{\downarrow}$ state to the presence or absence of the corresponding atom.

\subsection{Addressing beams}\label{SubSM:addressing}

The addressing laser beams are generated by an external cavity, $1013~$nm diode laser 
seeding an amplifier outputting up to $8~$W. The light is blue detuned from the 
$(6P_{3/2}, m_J=3/2)$ to $\ket{\uparrow}$ transition by $\Delta/2\pi \sim 400~$MHz. 
We use a second dedicated SLM to produce the pattern 
of addressing beams, superimposed onto the tweezer array pattern. 
Each beam is focused on a $1/e^2$ radius of $1.5~\mu$m 
and induces a light-shift of either $1\delta$ or two $2\delta$. A power of $\approx 300$\, 
mW on one atom results into a light-shift $\delta/(2\pi)\approx 23$\, MHz.

\begin{figure}
	\centering
	\includegraphics{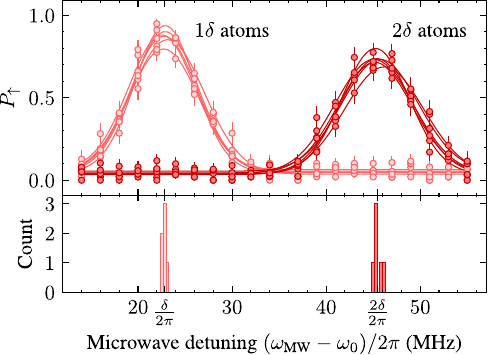}
	\caption{
	\textbf{Microwave spectroscopy to measure the light-shifts.}
	Top panel: probability of each addressed atom to be in $\ket{\uparrow}$ 
	after the microwave pulse at a frequency $\omega_{\text{MW}}$. 
	For each atoms, we fit the data by a Gaussian (thin curves) and extract the center frequency. 
	Bottom panel: histogram of the fitted frequencies. 
	The width of the bars represent the frequency fit uncertainty.}
	\label{fig:fig_SM_spectro}
\end{figure}

To control the light-shift on each addressed atoms, we perform a 
microwave spectroscopy on the $\uparrow-\downarrow$ transition: 
We initialize all the atoms in $\ket{\downarrow}$, turn on the addressing light, 
send a microwave pulse at a given frequency near $\omega_0$ and readout the state. 
Figure.~\ref{fig:fig_SM_spectro} shows a typical spectrum obtained after optimization 
of the SLM parameters to apply the desired light-shifts. 
The average light-shifts for the $1\delta$ (and $2\delta$) atoms are 
$2\pi \times 22.82~$MHz ($2\pi \times 45.46~$MHz) with a typical dispersion of 
$2\pi \times 0.2~$MHz ($2\pi \times 0.4~$MHz) between the different atoms. 

A current limitation of this scheme is that the only characteristic of the addressing pattern that can be dynamically modified during the $\sim 10\,\mu$s timescale of an experiment is the overall amplitude of the light-shift; however, one could envision to circumvent this limitation in the future, using e.g. several SLMs in a multiplexing configuration.

\subsection{Calibration of the phase $\phi(t_{\text{phase}})$}\label{SubSM:phase}

\begin{figure}
	\centering
	\includegraphics{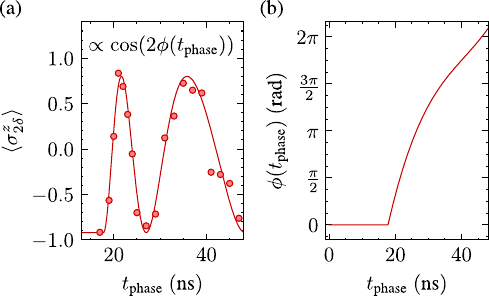}
	\caption{
	\textbf{Calibration of $\phi(t_{\text{phase}})$.}
	(a)  2$\delta$ atoms magnetization as a function of $t_{\text{phase}}$ (circles). 
	Solid curve: results obtained after fitting by a sinusoidal function 
	with a phase  $\phi(t_{\text{phase}})$ varying as piecewise polynomial function.
	(b) Fitted phase as a function of $t_{\text{phase}}$.}
	\label{fig:fig_SM_phase}
\end{figure}

In order calibrate the imprinted phase $\phi$ as a function of $t_{\text{phase}}$ 
we perform a Ramsey experiment. The atoms are first initialized in $\ket{\uparrow}$. 
We apply successively a $\pi/2$ microwave pulse, send the addressing pulse for 
a time $t_{\text{phase}}$, reapply a $\pi/2$ microwave pulse and readout in the $z$-basis. 
Figure.~\ref{fig:fig_SM_phase} shows the $2\delta$-atom magnetization as a function of 
$t_{\text{phase}}$. We expect a magnetization oscillation at a frequency $2\delta$ 
but due to transient effects in the EOM, the accumulated phase $\phi(t_{\text{phase}})$ 
exhibits a non-linear behavior with time. 
We use a piecewise polynomial function to fit $\phi(t_{\text{phase}})$ 
(solid curves in Fig.~\ref{fig:fig_SM_phase}). 
This fit allows us to convert $t_{\text{phase}}$ into $\phi$.

\section{Experimental imperfections}\label{SM:Exp_imperf}

In this section, we review the sources of errors decreasing 
the preparation and detection fidelities of the  3-atom states studied in the main text.

\subsection{State preparation errors}\label{SubSM:State_prep_imperf}

We estimate the fidelity of the Rydberg excitation $\eta_{\text{STIRAP}} = 98\pm0.3\%$ \cite{Chen2023,Bornet2023}. 
A fraction $1-\eta_{\text{STIRAP}}$ of the atoms stays in $5S_{1/2}$ after the STIRAP sequence 
and hence do not participate in the dynamics. 
These non-interacting atoms are read as a spin $\ket{\uparrow}$ at the end of the sequence.

\subsection{Local rotations imperfections}\label{SubSM:Rotations_imperf}

We identified six sources of imperfections occurring during the local rotations sequence 
shown in Fig.~\ref{fig:fig_SM_sequence}). \\

{\it Finite light-shifts} --  The applied light-shift of $\delta/2\pi \approx 23~$MHz are 
not much larger compared to the microwave Rabi frequency $\Omega_{\max}/2\pi = 5.43~$MHz. 
This induces crosstalk with the off resonant microwave leading to imperfect rotations of the atoms.

{\it Interaction during rotations} -- Since the XY interactions cannot be switched off,
the atoms interact during the $\sim100~$ns of local rotations. 
These residual interactions affect the rotations and thus reduce the detection fidelity. Experimentally, we optimized the microwaves Rabi frequencies $\Omega_{\max}$ to minimize this local rotation duration time while keeping $\delta\gg \Omega_{\max}$ to maximize the rotation efficiencies.

{\it Addressing-induced depumping} -- Due to the spontaneous emission 
induced by off-resonant coupling to the short-lived intermediate state $6P_{3/2}$, 
the addressed atoms in $\ket{\uparrow}$ are slowly depumped to the ground state $5S_{1/2}$. 
For $\Delta/2\pi \sim 400~$MHz and $\delta/2\pi \approx 23~$MHz we experimentally measure 
effective lifetimes of $\sim 2.3~\mu$s and $\sim 1.1~\mu$s for 
the $1\delta$ and $2\delta$ atoms in the $\ket{\uparrow}$ state.

{\it Addressing-induced atom losses} -- The tightly focused addressing beams apply a ponderomotive force 
on the addressed atoms, pushing them away from their trap center, thus preventing them from being recaptured
before readout. Experimentally, for $\delta/2\pi \approx 23$\, MHz, we measure losses of $0.3\pm0.3\%$ and $1.3\pm0.3\%$ 
for the $1\delta$ and $2\delta$ atoms when sending a $80~$ns addressing pulse.

{\it Light-shifts inhomogeneities} -- As explained in a previous section,
the light-shifts applied to the atoms are not perfectly homogeneous. 
We measured a dispersion on the order of $1\%$ after calibration that can drift up to $3\%$ 
after one day without further calibrations. 
When the addressing is on, the dispersion results into a variation of the phase accumulation of the $1\delta$-atoms 
across the array. This leads to a spread of the angle of rotation of the qubits 
when sending the microwave pulses for local rotations. 

{\it Timing jitter} --  We measure an electronic jitter of $\pm 2$\,ns between the addressing 
and the microwave pulses. It has a similar effect to that of light shift inhomogeneities. 
Shot-to-shot, the jitter induces an uncertainty in the angle of rotation of the $1\delta$ atoms leading to imperfect microwave rotations.

\subsection{Readout imperfections}\label{SubSM:Readout_imperf}

Due to the finite efficiency of each step in the readout sequence shown in Fig.~\ref{fig:fig_SM_sequence}, 
an atom in  $\ket{\uparrow}$ (respectively in $\ket{\downarrow}$) has a small probability 
$\epsilon_{\uparrow}$ ($\epsilon_{\downarrow}$) to be detected in 
$\ket{\downarrow}$ ($\ket{\uparrow}$). The main contributions to $\epsilon_{\uparrow}$ are
the finite efficiency of the deexcitation pulse $\eta_{\text{deexc}}$ and the probability 
$\epsilon$ to lose an atom due to background gas collisions. 
For $\epsilon_{\downarrow}$ the main contribution is the radiative decay 
of $\downarrow$ to the ground state. A set of independent experiments 
leads to first order to  
$\epsilon_{\uparrow} = (1-\eta_{\text{deexc}}) + \epsilon = 1.5\% + 1.2\% = 2.7 \pm 0.3\%$ 
and $\epsilon_{\downarrow} = 1.5\pm0.3\%$.

Using these values, we correct the data of the quantum state tomography from detection errors. 
We denote ${\bm P^i}=(P_{\uparrow}^i, P_{\downarrow}^i)$ 
the probability to measure atom $i$ in $\ket{\uparrow}$ and $\ket{\downarrow}$. 
These raw probalities are linked to their corresponding quantities 
$ \tilde{{\bm P}}^i=(\tilde{P}_{\uparrow}^i, \tilde{P}_{\downarrow}^i)$ 
free from detection errors by ${\bm P^i} = M\tilde{{\bm P}}^i$, with $M$ the detection error matrix:
\begin{equation}\label{Eq:correction}
M = 
\begin{pmatrix}
	1-\epsilon_{\uparrow} & \epsilon_{\downarrow} \\
	\epsilon_{\uparrow} & 1-\epsilon_{\downarrow}
\end{pmatrix}
.
\end{equation}
Extending to 3-atoms and assuming uncorrelated errors, we get 
${\bm P^i} \otimes {\bm P^j} \otimes {\bm P^k} = 
(M \otimes M \otimes M)(\tilde{{\bm P}}^i \otimes \tilde{{\bm P}}^j \otimes \tilde{{\bm P}}^k)$. 
To correct the raw data for detection errors we invert $(M \otimes \cdots \otimes M)$ 
to compute $(\tilde{{\bm P}}^i \otimes \cdots \otimes \tilde{{\bm P}}^k)$.

\subsection{Decoherence during the dynamics}\label{SubSM:Decoherence}

Besides the state preparation and measurement errors described above, 
two other sources of errors contribute to decoherence.\\

{\it Positional disorder} -- Due to the extension of the wave-packet in the traps and the residual atomic motion, 
the interatomic distances exhibit shot-to-shot fluctuations. We consider that the atomic positions and velocities 
follow a Gaussian distribution with standard deviations 
$\sigma_{(x,y)} \sim 100~$nm in the array plane and $\sigma_{z} \sim 0.6~\mu$m perpendicular to it, and 
${\bm \sigma}_v\sim 0.03~\mu\text{m}/\mu$s in all directions. 
This positional disorder leads to shot-to-shot fluctuations of the interaction energy 
between two atoms and thus results in decoherence 
when averaging over experimental realizations.

{\it Finite Rydberg lifetime} -- During the dynamics, one atom in a Rydberg state can decay 
spontaneously to the ground state (rate $\Gamma^{\text{sp}}$) or be transferred by black-body 
radiation to other Rydberg states (rate $\Gamma^{\text{bb}}$). 
When this occurs, the readout of the qubit state can be biased. 
Moreover, a decay induces a loss from the qubit basis
and thus affects the ensuing dynamics. 
We calculate from~\cite{Beterov2009} for $T = 300~$K: 
$1/\Gamma^{\text{sp}}_{60S} \simeq 260~\mu$s,  
$1/\Gamma^{\text{bb}}_{60S} \simeq 157~\mu$s, 
$1/\Gamma^{\text{sp}}_{60P} \simeq 472~\mu$s and 
$1/\Gamma^{\text{bb}}_{60P} \simeq 161~\mu$s.

\section{Numerical simulations}\label{SM:Simulations}

The simulations used to benchmark the 3-atom results of Fig.\,1 and 2 
include the processes described in the previous section. 
Each atom is modeled as a 5-level system with the following basis states: 
the two interacting spin states $\ket{\uparrow}$ and $\ket{\downarrow}$, 
the ground-state $\ket{g}$ where atoms are initialized and can decay to from all the others states, 
the short-lived intermediate state $6P_{3/2}$ used in the two-photon excitation,
and $\ket{r}$, effectively representing the Rydberg manifold other than $\ket{\uparrow}$ and $\ket{\downarrow}$. 
The whole sequence depicted in Fig.~\ref{fig:fig_SM_sequence} is implemented with timings, 
amplitudes and frequencies as experimentally realized and including  
all decay processes mentioned in the previous sections. 
The only exception is the freezing protocol, which is modeled by a fast decay process
from $\ket{\downarrow}$ to $\ket{r}$. 
The qubit states interact both via the XY Hamiltonian with coupling $J/h = -0.82$\, MHz,
and by van der Waals interaction with $C_6^{\uparrow\uparrow}/a^6\approx 2\pi\times 40$\,kHz  
and $C_6^{\downarrow\downarrow}/a^6\approx 2\pi\times 6$\,kHz. 
We simulate the dynamics of this $5^3$-level system using the package \textit{qutip}, 
starting from $\ket{g}^{\otimes 3}$ and solving the master equation for each step 
of the sequence and each set of experimental parameters.

We account for the shot-to-shot fluctuations in the interatomic distances and timing jitters by a 
Monte-Carlo sampling of the experimental parameters with the estimated bounds given 
in the previous sections. We also include inhomogeneities of the light-shifts 
across the array. As solving the Master equation becomes quickly resource-consuming 
($\sim1$ hour/shot), we average the simulation results over only $20$ shots, 
which appears to be enough to reproduce the experimental data with reasonable agreement. 
At the end of the Monte Carlo simulation, we apply to the results the addressing-induced 
atom losses and readout errors following the procedure described in previous the sections.
The simulation predicts a STIRAP efficiency for one atom of $98.3\%$, in very good agreement with the experiment.
The phases of the microwave pulses are optimized so as to reproduce the data in Fig.~1(c), and are 
then fixed to these values.

To simulate the chirality curves shown in Fig.~2(c) and assess the impact of 
the various imperfections, we first calculate the preparation fidelity of the W-state. 
We obtain $\eta_{W}=83\pm0.4\%$, with   contributions to the infidelity of 6\% from the 
STIRAP finite efficiency, 6\% from the Rydberg lifetime and 5\% from positional disorder.
We then include the phase imprinting step to prepare the state $\ket{\chi(\phi)}$. 
Calculating the chirality $S$ of this state yields a maximum (minimum) value of 
$S/S_{\rm max}=0.8 (-0.79)$ (red curve in Fig.~2(c)) with $S_{\max} = 2\sqrt{3}$. 
Finally we simulate the measurement phase for each of the 6 components of the chirality 
and obtain maximum (minimum) values $S/S_{\rm max}=0.55 (-0.44)$ (purple curve), 
in good agreement with the data. Applying the measurement sequence starting from 
a perfect $\ket{\chi(\phi)}$ leads to $S/S_{\rm max}=0.7 (-0.62)$ (blue curves).

\section{Quantum state tomography}\label{SM:Tomography}

To perform the full tomography of the 3-atom $\ket{W}$ and $\ket{\chi^\pm}$ states, we measure
the 27 observables summarized in the first column of Tab.~\ref{Tab:Tab_rotations_MW}. \\

\begin{center}	
	\begin{table}
		\begin{tabular}{|c|c|c|c|c|}
			\hline Basis ($0\delta$,$1\delta$,$2\delta$) & Rotations &  $\phi_{\text{all}}($\degree$)$ &  $\phi_{0\delta}($\degree$)$ &  $\phi_{1\delta}($\degree$)$  \\
			\hline $xxx$ & $R^{-y}\piot$ 															& $-90$ 	& $-$ 		& $-$  \\
			\hline $xxy$ & $\left(R_{0\delta}^{-y} \otimes R_{1\delta}^{-y}\piot\right) \cdot R^{x}\piot$		& $0$ 	& $-90$	& $-90$  \\
			\hline $xxz$ & $R_{0\delta}^{-y}\piot \otimes R_{1\delta}^{-y}\piot$							& $-$ 	& $-90$ 	& $-90$  \\
			\hline $xyx$ & $R_{1\delta}^{x}\piot \cdot R^{-y}\piot$							 			& $-90$ 	& $-$ 	& $0$  \\
			\hline $xyy$ & $R_{0\delta}^{-y}\piot \cdot R^{x}\piot$ 										& $0$ 	& $-90$ 	& $-$  \\
			\hline $xyz$ & $R_{0\delta}^{-y}\piot \otimes R_{1\delta}^{x}\piot$ 							& $-$ 	& $-90$ 	& $0$  \\
			\hline $xzx$ & $R_{1\delta}^{y}\piot \cdot R^{-y}\piot$ 										& $-90$ 	& $-$ 	& $90$  \\
			\hline $xzy$ & $\left(R_{0\delta}^{-y}\piot \otimes R_{1\delta}^{-x}\piot\right)  \cdot R^{x}\piot$	& $0$ 	& $-90$ 	& $180$  \\
			\hline $xzz$ & $R_{0\delta}^{-y}\piot$ 													& $-$ 	& $-90$ 	& $-$  \\
			\hline $yxx$ & $R_{0\delta}^{x}\piot \cdot R^{-y}\piot$ 										& $-90$ 	& $0$ 	& $-$  \\
			\hline $yxy$ & $R_{1\delta}^{-y}\piot \cdot R^{x}\piot$ 										& $0$ 	& $-$ 	& $-90$  \\
			\hline $yxz$ & $R_{0\delta}^{x}\piot \otimes R_{1\delta}^{-y}\piot$	 						& $-$ 	& $0$ 	& $-90$  \\
			\hline $yyx$ & $\left(R_{0\delta}^{x}\piot \otimes R_{1\delta}^{x}\piot\right)  \cdot R^{-y}\piot$		& $-90$ 	& $0$ 	& $0$  \\
			\hline $yyy$ & $R^{x}\piot$ 															& $0$ 	& $-$ 	& $-$  \\
			\hline $yyz$ & $R_{0\delta}^{x}\piot \otimes R_{1\delta}^{x}\piot$ 							& $-$ 	& $0$ 	& $0$  \\
			\hline $yzx$ & $\left(R_{0\delta}^{x}\piot \otimes R_{1\delta}^{y}\piot\right)  \cdot R^{-y}\piot$ 	& $-90$ 	& $0$ 	& $90$  \\
			\hline $yzy$ & $R_{1\delta}^{-x}\piot \cdot R^{x}\piot$ 										& $0$ 	& $-$ 	& $180$  \\
			\hline $yzz$ & $R_{0\delta}^{x}\piot$ 													& $-$ 	& $0$ 	& $-$  \\
			\hline $zxx$ & $R_{0\delta}^{y}\piot \cdot R^{-y}\piot$ 										& $-90$ 	& $90$ 	& $-$  \\
			\hline $zxy$ & $\left(R_{0\delta}^{-x}\piot \otimes R_{1\delta}^{-y}\piot\right)  \cdot R^{x}\piot$	& $0$ 	& $180$ 	& $-90$  \\
			\hline $zxz$ & $R_{1\delta}^{-y}\piot$ 													& $-$ 	& $-$ 	& $-90$  \\
			\hline $zyx$ & $\left(R_{0\delta}^{y}\piot \otimes R_{1\delta}^{x}\piot\right)  \cdot R^{-y}\piot$		& $-90$ 	& $90$ 	& $0$  \\
			\hline $zyy$ & $R_{0\delta}^{-x}\piot \cdot R^{x}\piot$ 										& $0$ 	& $180$ 	& $-$  \\
			\hline $zyz$ & $R_{1\delta}^{x}\piot$ 													& $-$ 	& $-$ 	& $0$  \\
			\hline $zzx$ & $\left(R_{0\delta}^{y}\piot \otimes R_{1\delta}^{y}\piot\right)  \cdot R^{-y}\piot$ 	& $-90$ 	& $90$ 	& $90$  \\
			\hline $zzy$ & $\left(R_{0\delta}^{-x}\piot \otimes R_{1\delta}^{-x}\piot\right)  \cdot R^{x}\piot$	& $0$ 	& $180$ 	& $180$  \\
			\hline $zzz$ & $\mathbb{1}$ & $-$ 	& $-$ 	& $-$  \\
			\hline 
		\end{tabular}
		\caption{
		\textbf{Microwave pulse sequence for the state tomography.} 
		First column:  measurement basis for the $0\delta$, $1\delta$ and $2\delta$ atoms. 
		Second column:  applied rotations. 
		Three last columns:  relative phases of the microwave pulses used implement the corresponding 
		rotations ($\phi_{\text{all}}$ refers to the global rotation and 
		$\phi_{0,1\delta}$ to the local ones). The $-$ symbol indicates that the 
		corresponding pulse is off for this sequence.}
	\label{Tab:Tab_rotations_MW}
	\end{table}
\end{center}

In order to reconstruct their density matrices $\rho$, we perform a maximum-likelihood estimation 
to constrain $\rho$ to be physical. We follow the method described in 
the Supplemental Materiel of~\cite{Takeda2021}. 
Any density matrix can be written as $\rho(T) = T^{\dag}T/\text{Tr}(T^{\dag}T)$ with $T$ 
being a complex $8\times8$ lower triangular matrix with real diagonal elements. 
Thus $T$ has 64 independent real parameters $(t_1,t_2,\cdots,t_{64})$ 
that minimize the following cost function:
\begin{equation}\label{Eq:cost_func}
	C(T) = \sum_{\alpha \in \{ x,y,z \}^{3}}\sum_{\beta \in \{ \uparrow,\downarrow \}^{3}}  \left(  \expval{R^{\dag}_\alpha \rho(T) R_\alpha}{\beta }-P_\alpha^\beta    \right)^2.\nonumber
\end{equation}
Here, $\alpha$ is the basis in which we measure each atom, 
$\beta$ is an experimental outcome,  
$P_\alpha^\beta$ the probability to measure $\beta$ in the $\alpha$ basis and 
$R_\alpha$ the set of applied rotations to measure in $\alpha$. 
For example, when measuring in the $xyz$-basis, 
$R_{xyz} = R_{0\delta}^{-y} \otimes R_{1\delta}^{x} \otimes \mathbb{1}_{2\delta}$. 
We perform the minimization using the L-BFGS-B algorithm provided by the SciPy package of Python.

\section{6-body chiral-chiral correlation function}

In this section we present additional numerical evidence demonstrating that $\langle \chi_A\chi_B\rangle'$ accurately tracks the dynamics of $\langle \chi_A\chi_B\rangle$ and thus serves as a suitable probe of the 6-body chiral-chiral correlations of the underlying spin system.
The full chiral-chiral correlation function is given by:
\begin{align}
    \langle \chi_A\chi_B\rangle &= \hspace{-5mm}\sum_{a,b,c \in \{x,y,z\}}
    \sum_{d,e,f  \in  \{x,y,z\}}
    \text{sgn}(a,b,c)\text{sgn}(d,e,f)\times \notag \\ 
    &\times \left[ \langle \sigma^a_1\sigma^b_2\sigma^c_3 \sigma^d_4\sigma^e_5 \sigma^f_6\rangle - \langle \sigma^a_1\sigma^b_2\sigma^c_3 \rangle \langle\sigma^d_4\sigma^e_5 \sigma^f_6\rangle \right]
\end{align}
where $\text{sgn}(a,b,c)$ corresponds to the parity of the permutation $(a,b,c)$ and $1-3$ and $4-6$ labels the spins of triangles $A$ and $B$ counter-clockwise, respectively.

When considering $\langle \chi_A\chi_B\rangle'$, the above sum is restricted only to the terms where the same measurement basis occurs for atoms of the same group.
As a result, $\text{sgn}(a,b,c)\text{sgn}(d,e,f)$ becomes a constant whose value depends on the relative handedness of the addressing light pattern between the two triangles.
In Eq.2, this phase factor is $\eta$, taking the value of $\eta=-1$ for Pattern 1 and $\eta=1$ for pattern 2 [Fig.~4(a)].

\begin{figure}[t]
    \centering
    \includegraphics[width = 3.2in]{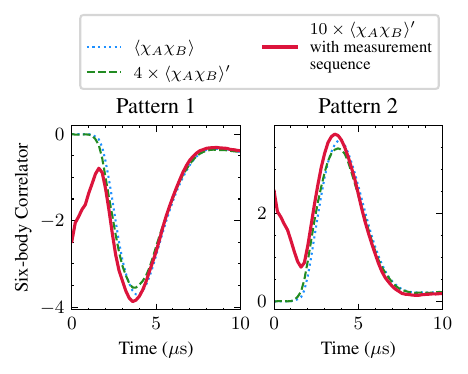}  
    \caption{Comparison of the full chiral-chiral correlation function $\langle \chi_A\chi_B\rangle$ and measured one $\langle \chi_A \chi_B\rangle'$ with and without measurement error (which includes both the full measurement sequence as well as detection errors).}
    \label{fig:fig_SM_agreement}
\end{figure}

Turning to the dynamics of the chiral-chiral correlations, we observed that, up to rescaling, the different correlation functions (full $\langle \chi_A\chi_B\rangle$, $\langle \chi_A\chi_B\rangle'$ and $\langle \chi_A\chi_B\rangle'$ with measurement imperfections) all follow one another at late times when the system is in either the ground or first excited state of the target Hamiltonian, Fig.~\ref{fig:fig_SM_agreement}. For short times $t\lesssim 1.5\,\mu$s, we observe non-zero partial chiral-chiral correlation when simulating the measurement imperfections. We attribute this effect to a waiting time before the measurements. We use an acoustic-optic modulator (AOM) to ramp down the addressing light intensity adiabatically. We need to set the AOM at maximum power to perform the local rotations. This step takes $150\,$ns. We switch off the addressing light using the EOM during this waiting time. When measuring for relatively early times of the ramp, the system is quenched far from equilibrium and thus freely evolves for $150\,$ns. More particularly, when quenched, the initial product state $\ket{\uparrow\downarrow\cdots}$ rapidly builds up chiral-chiral correlations that we measure.

\section{Diabatic Errors in Adiabatic Protocol}

\begin{figure}[t]
    \centering
    \includegraphics[width = 3.4in]{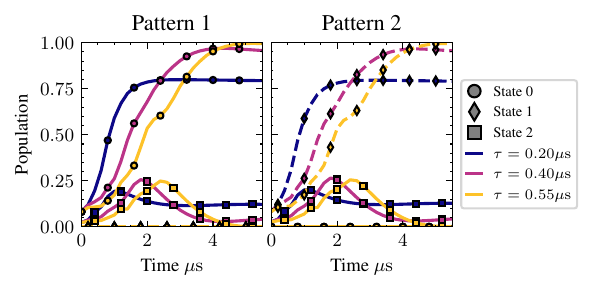}
    \caption{
      Effect of varying $\tau$ in the adiabatic protocol patterns 1 and 2.
      When $\tau\gtrsim 0.55\mathrm{\mu s}$, the population on the final correct state is above $99\%$.
      Simulations were performed without taking into account imperfections in the experiment. }
    \label{fig:fig_SM_adiabaticity}
\end{figure}

While the adiabatic protocol offers a powerful tool for the preparation of an eigenstate, the finite duration of the adiabatic ramp will inevitably lead to a small, but finite error.
In this section, we analyze the effect of the duration of the adiabatic protocol (as characterized by the timescale $\tau$ of the light-shift decay) on success probability of the adiabatic protocol [Fig.~\ref{fig:fig_SM_adiabaticity}].
We find that by choosing $\tau\gtrsim 0.55~\mathrm{\mu s}$, the population in the target final state is above $99\%$ and thus our protocol is not limited by these diabatic errors.

While these diabatic errors do not affect significantly the late time population, they can have important dynamical features.
Throughout the adiabatic protocols considered, the instantaneous Hamiltonian is time reversal symmetry and, as a result, the instantaneous eigenstates will not exhibit a non-zero chirality.
Diabatic errors break time-reversal symmetry and can generate coherent superpositions of eigenstates that can exhibit non-zero chiralities.
Indeed, this is what we observe [Fig.~\ref{fig:fig_SM_Imperfections}(a)].

Interestingly, the frequency of the oscillation provides a direct probe of the system's energy levels.
Since the population is mainly restricted to two levels, the frequency of the oscillations reflects the energy difference between the ground and the second excited state [black line in Fig.~\ref{fig:fig_SM_Imperfections}(a2)]

\section{Effect of Experimental Imperfections in Adiabatic Protocol}

\begin{figure}[t]
	\centering
	\includegraphics[width = 3.0in]{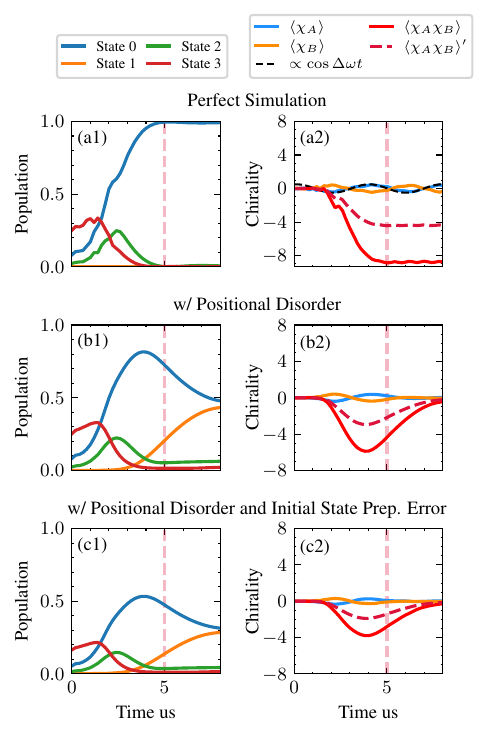}
	\caption{Dynamics of the adiabatic protocol for pattern 1 with $\tau=0.55\mathrm{\mu s}$.
		We compare three different scenarios: the perfect adiabatic protocol (a), the adiabatic protocol with positional disorder (b), and the adiabatic protocol with positional disorder and initial state preparation error (c).
		The positional disorder induces a shot-to-shot variation in the interactions of the system that is ultimately responsible for the equilibration of the populations between the two lowest energy state (State 0 and 1), causing the decay of the chiral-chiral correlations function.
		By contrast, initial state preparation errors result in an overall attenuation of the signal.
	}
	\label{fig:fig_SM_Imperfections}
\end{figure}

In this section we illustrate how other imperfections in the experiment (namely, positional disorder and initial state preparation error) alter our adiabatic protocol.
We focus on the case of pattern 1, incorporating the two effects at a time [Fig.~\ref{fig:fig_SM_Imperfections}].

As discussed in a previous appendix, positional disorder can be modeled as a shot-to-shot variation of the position and velocities of the different atoms.
This change in positions implies a change in the interactions between the different atoms.
As a result, each realization of the experiment will prepare a slightly different state which will differ from the precise target state of interest.

Dynamically, this manifests itself as a equilibration of the populations between the ground and first excited states, the two lowest lying states that are energetically separated from the remaining eigenstates of the system.
This can be observed in the dynamics of the population of the perfect initial state under the adiabatic protocol, Fig.~\ref{fig:fig_SM_Imperfections}(b1).
At early times, the dynamics of the populations match the disorder-free case [Fig.~\ref{fig:fig_SM_Imperfections}(a1)], until around $t\sim 4\mathrm{\mu s}$, where the population of the other low energy state (orange) suddenly increases until it approximately matches the population of the target state (blue).
This explains the experimental observation of the peak in the chiral-chiral correlations---the subsequent decay does not arise from decoherence of the atoms, but rather the dynamical effect of the positional disorder~[Fig.~\ref{fig:fig_SM_Imperfections}(b2)].

Finally, we consider also the effect of initial state preparation error.
In this case, when preparing the wrong initial state, we no longer have guarantees of preparing a state close to the target state.
As a result, we observe an overall decrease in the signal that is consistent with the probability of preparing the correct initial state for our 6-atom cluster ($\approx 63\%$).
Besides this drop in the signal, all other dynamical features remain the same.

\end{document}